\documentclass[a4paper,USenglish]{lipics-v2021}

\pdfoutput=1
\hideLIPIcs
\nolinenumbers

\usepackage{float}
\usepackage{booktabs}

\title{SATisfying the High School Identities but not Wilkie's Identity}

\author{Agon Hajdari}{%
  Department of Computer Science, University of Innsbruck, Innsbruck, Austria%
}{agonmhajdari@gmail.com}{}{}

\author{Johannes Niederhauser}{%
  Department of Computer Science, University of Innsbruck, Innsbruck, Austria%
}{johannes.niederhauser@uibk.ac.at}{https://orcid.org/0000-0002-8662-6834}{}

\authorrunning{A.~Hajdari and J.~Niederhauser}



\supplement{\url{https://doi.org/10.5281/zenodo.18568303}}

\begin{document}

\maketitle

\begin{abstract}
We settle an open question related to Tarski’s High School Algebra problem by showing
that no 11-element algebra can satisfy the High School Identities while refuting
Wilkie’s identity. We encode the search as a SAT instance and independently verify
the result. As a byproduct, we obtain a new 12-element countermodel that is not
isomorphic to the previously known one.
\end{abstract}

\section{Introduction}
\label{sec:introduction}
Tarski’s High School Algebra problem asks whether every identity about
addition, multiplication and exponentiation which holds over the natural numbers is derivable from
the High School Identities (HSI). Wilkie answered this negatively by exhibiting
an identity true in $\mathbb{N}$ but not derivable from HSI~\cite{burris2005saga}.
A standard way to show non-derivability is to build a model of HSI that
falsifies the identity. Therefore, Much of the subsequent work focused on
constructing small countermodels and narrowing the minimum size. By 2005 a
12-element countermodel was known, and exhaustive searches ruled out sizes
up to 10, leaving the 11-element case open~\cite{burris2005saga,zhang2005computer}.

In this paper, we formulate the search for finite countermodels as a SAT problem
and solve the resulting instances with parallel SAT solvers.
We give a direct CNF encoding of HSI together with the negation of Wilkie’s
identity, augmented by symmetry breaking and pruning lemmas from the algebraic
literature. Using proof-producing solvers and independently verified certificates,
we show that all instances of size $n \le 11$ are unsatisfiable, settling the last
open size below 12. As a byproduct, we obtain a distinct 12-element countermodel
that is not isomorphic to the previously known one.

As has already been noted by Zhang
more than 20 years ago in the paper performing exhaustive searches for countermodels
up to size 10, the problem of finding countermodels to Wilkie's identity is highly equational
and therefore relatively hard for
SAT solvers \cite{zhang2005computer}. Back then, SAT solvers could not be used as
a main tool for the exhaustive search, but our results show that modern parallel
SAT solvers are up to the task even if we go for $n = 11$. Hence, resolving this open problem using SAT is yet another
indicator of the progress which the community has made in recent years. Our
paper is structured as follows:
Section~\ref{sec:algebra} recalls the algebraic background and the structural
constraints on countermodels. Section~\ref{sec:sat} presents the SAT encoding.
Finally, Section~\ref{sec:experiments} reports on solver performance and proof verification
before we conclude in Section~\ref{sec:conclusion}.

\section{Algebraic Background}
\label{sec:algebra}

This section recalls the algebraic setting behind Tarski's High School Algebra
problem and Wilkie's identity, and summarizes the structural facts that
motivate the SAT encoding in Section~\ref{sec:sat}.
An \emph{equational theory} is a set of universally quantified equations over a
signature; here the signature consists of the binary operations $+$, $\cdot$,
$\operatorname{exp}$ and a constant $1$.
An equation $s=t$ is \emph{valid} in an algebra $\mathcal{A}$ if it holds under
all variable assignments in $\mathcal{A}$, and \emph{derivable} from axioms if
it follows by equational reasoning
(reflexivity, symmetry, transitivity, substitution and congruence).
For first-order equational theories, Birkhoff showed that
derivability and validity in
all models of the axioms coincide~\cite{birkhoff1935}. However, there
may be models of the axioms which enjoy equalities which are not derivable
from the axioms.

The \emph{High School Identities} (HSI) are a finite collection of equations
capturing the familiar interaction of addition, multiplication, and
exponentiation; they are surveyed in Burris and Yeats~\cite{burris2005saga}.
The axioms are, for all variables,
\begin{align*}
x + y &= y + x, &
(x + y) + z &= x + (y + z), &
x \cdot (y + z) &= (x \cdot y) + (x \cdot z), \\
x \cdot y &= y \cdot x, &
(x \cdot y) \cdot z &= x \cdot (y \cdot z), &
x \cdot 1 &= x, \\
x^{y+z} &= x^y \cdot x^z, &
(x \cdot y)^z &= x^z \cdot y^z, &
(x^y)^z &= x^{y \cdot z}, \\
1^x &= 1, &
x^1 &= x.
\end{align*}
These identities hold in the standard structure
$(\mathbb{N}, +, \cdot, \operatorname{exp})$.
They generate the equational theory $\mathsf{HSI}_{\exp}$ over
$\{+, \cdot, \operatorname{exp}, 1\}$.
Tarski asked whether the high school identities are complete for arithmetic
exponentiation, i.e., whether every identity valid in $\mathbb{N}$ is
derivable from HSI. Wilkie exhibited an identity
\[
(P^x + Q^x)^y \cdot (R^y + S^y)^x
\;=\;
(P^y + Q^y)^x \cdot (R^x + S^x)^y,
\]
with
\begin{align*}
P &= 1 + x, &
Q &= 1 + x + x^2, &
R &= 1 + x^3, &
S &= 1 + x^2 + x^4,
\end{align*}
that holds in $\mathbb{N}$ but is not derivable from HSI~\cite{burris2005saga}.
Non-derivability can be witnessed semantically: It suffices to build a model
satisfying all HSI axioms while violating the identity.

A \emph{HSI-algebra} is a structure
$\mathcal{A}=(A,+,\cdot,\operatorname{exp},1)$ satisfying HSI.
A finite HSI-algebra is a \emph{countermodel} to Wilkie's identity if it
satisfies HSI but falsifies the identity; following Zhang, such countermodels
are called \emph{Gurevi\v{c}--Burris algebras (GBA)}~\cite{zhang2005computer}.
Gurevi\v{c} constructed a 59-element countermodel, and subsequent work reduced the
bound; by 2005, a 12-element countermodel was known while no example of size at
most 10 had been found, leaving the 11-element case as the critical
gap.
Historically, explicit countermodels were found with sizes 59 (Gurevi\v{c}, 1985),
28 (Burris, 1988), 16 (Burris, 1990), 15 (Lee, 1991), 14 (Jackson, 1996),
13 (Burris--Yeats, 2001), and 12 (Burris--Yeats, 2001), which remains the
smallest known example~\cite{burris2005saga}. Lower bounds advanced as well:
Burris--Lee \cite{burris1992} ruled out sizes below 7 (1992), Jackson \cite{jackson1996} ruled out sizes below 8
(1996), and Zhang \cite{zhang2005computer} (2005) ruled out sizes at most 10 by exhaustive search
using SEM~\cite{zhang1995sem} and Mace4~\cite{mccune2003mace4}.
Table~\ref{tab:operations-12} shows a new 12-element countermodel found by us which
is not isomorphic to the one reported in~\cite{burris2005saga}. We have more to say
about in in Section~\ref{sec:experiments}.

\begin{table}[t]
\centering
\scriptsize
\setlength{\tabcolsep}{3pt}
\renewcommand{\arraystretch}{1.2}
\begin{tabular}{c|*{12}{r}}
$+$ & 1 & 2 & 3 & 4 & 5 & 6 & 7 & 8 & 9 & 10 & 11 & 12 \\
\midrule
1 & 2 & 3 & 11 & 2 & 3 & 9 & 11 & 11 & 3 & 3 & 11 & 11 \\
2 & 3 & 11 & 11 & 3 & 11 & 3 & 11 & 11 & 11 & 11 & 11 & 11 \\
3 & 11 & 11 & 11 & 11 & 11 & 11 & 11 & 11 & 11 & 11 & 11 & 11 \\
4 & 2 & 3 & 11 & 10 & 12 & 10 & 11 & 11 & 3 & 11 & 11 & 11 \\
5 & 3 & 11 & 11 & 12 & 11 & 11 & 3 & 12 & 11 & 11 & 11 & 11 \\
6 & 9 & 3 & 11 & 10 & 11 & 10 & 11 & 11 & 3 & 11 & 11 & 11 \\
7 & 11 & 11 & 11 & 11 & 3 & 11 & 11 & 3 & 11 & 11 & 11 & 11 \\
8 & 11 & 11 & 11 & 11 & 12 & 11 & 3 & 11 & 11 & 11 & 11 & 11 \\
9 & 3 & 11 & 11 & 3 & 11 & 3 & 11 & 11 & 11 & 11 & 11 & 11 \\
10 & 3 & 11 & 11 & 11 & 11 & 11 & 11 & 11 & 11 & 11 & 11 & 11 \\
11 & 11 & 11 & 11 & 11 & 11 & 11 & 11 & 11 & 11 & 11 & 11 & 11 \\
12 & 11 & 11 & 11 & 11 & 11 & 11 & 11 & 11 & 11 & 11 & 11 & 11 \\
\end{tabular}

\vspace{0.6em}
\begin{tabular}{c|*{12}{r}}
$\cdot$ & 1 & 2 & 3 & 4 & 5 & 6 & 7 & 8 & 9 & 10 & 11 & 12 \\
\midrule
1 & 1 & 2 & 3 & 4 & 5 & 6 & 7 & 8 & 9 & 10 & 11 & 12 \\
2 & 2 & 11 & 11 & 10 & 11 & 10 & 11 & 11 & 11 & 11 & 11 & 11 \\
3 & 3 & 11 & 11 & 11 & 11 & 11 & 11 & 11 & 11 & 11 & 11 & 11 \\
4 & 4 & 10 & 11 & 6 & 12 & 6 & 11 & 11 & 10 & 10 & 11 & 11 \\
5 & 5 & 11 & 11 & 12 & 11 & 11 & 11 & 12 & 11 & 11 & 11 & 11 \\
6 & 6 & 10 & 11 & 6 & 11 & 6 & 11 & 11 & 10 & 10 & 11 & 11 \\
7 & 7 & 11 & 11 & 11 & 11 & 11 & 11 & 11 & 11 & 11 & 11 & 11 \\
8 & 8 & 11 & 11 & 11 & 12 & 11 & 11 & 11 & 11 & 11 & 11 & 11 \\
9 & 9 & 11 & 11 & 10 & 11 & 10 & 11 & 11 & 11 & 11 & 11 & 11 \\
10 & 10 & 11 & 11 & 10 & 11 & 10 & 11 & 11 & 11 & 11 & 11 & 11 \\
11 & 11 & 11 & 11 & 11 & 11 & 11 & 11 & 11 & 11 & 11 & 11 & 11 \\
12 & 12 & 11 & 11 & 11 & 11 & 11 & 11 & 11 & 11 & 11 & 11 & 11 \\
\end{tabular}

\vspace{0.6em}
\begin{tabular}{c|*{12}{r}}
$\operatorname{exp}$ & 1 & 2 & 3 & 4 & 5 & 6 & 7 & 8 & 9 & 10 & 11 & 12 \\
\midrule
1 & 1 & 1 & 1 & 1 & 1 & 1 & 1 & 1 & 1 & 1 & 1 & 1 \\
2 & 2 & 11 & 11 & 7 & 7 & 11 & 11 & 11 & 11 & 11 & 11 & 11 \\
3 & 3 & 11 & 11 & 5 & 8 & 11 & 11 & 5 & 11 & 11 & 11 & 12 \\
4 & 4 & 6 & 6 & 6 & 6 & 6 & 6 & 6 & 6 & 6 & 6 & 6 \\
5 & 5 & 11 & 11 & 11 & 12 & 11 & 11 & 11 & 11 & 11 & 11 & 11 \\
6 & 6 & 6 & 6 & 6 & 6 & 6 & 6 & 6 & 6 & 6 & 6 & 6 \\
7 & 7 & 11 & 11 & 11 & 11 & 11 & 11 & 11 & 11 & 11 & 11 & 11 \\
8 & 8 & 11 & 11 & 12 & 11 & 11 & 11 & 12 & 11 & 11 & 11 & 11 \\
9 & 9 & 11 & 11 & 7 & 11 & 11 & 11 & 11 & 11 & 11 & 11 & 11 \\
10 & 10 & 11 & 11 & 11 & 11 & 11 & 11 & 11 & 11 & 11 & 11 & 11 \\
11 & 11 & 11 & 11 & 11 & 11 & 11 & 11 & 11 & 11 & 11 & 11 & 11 \\
12 & 12 & 11 & 11 & 11 & 11 & 11 & 11 & 11 & 11 & 11 & 11 & 11 \\
\end{tabular}
\caption{Operation tables for the new 12-element algebra.}
\label{tab:operations-12}
\end{table}

Finite countermodels satisfy strong structural constraints.
Burris and Lee showed that any countermodel must have at least
three distinct integers in the subalgebra $1, 1+1, 1+1+1, \dots$
whose elements are called \emph{integers}.
In particular, they showed that the term values of
$1$, $2:=1+1$, and $3:=2+1$ are fixed and distinct in every countermodel~\cite{burris1992}.
Hence we may, without loss of generality, rename these elements to
$1,2,3$, which motivates fixing them in the SAT encoding presented in next section.
Let $u \mid v$ abbreviate $\exists w\,(v=u\cdot w)$ and set
$P=1+x$, $Q=1+x+x^2$, $R=1+x^3$, $S=1+x^2+x^4$.
If any of the divisibility relations $P\mid Q$, $Q\mid P$, $R\mid S$, or
$S\mid R$ holds, then Wilkie's identity follows; hence any countermodel with a
witness pair $(a,b)$ (that is, $W(a,b)$ false) must satisfy the constraints
\[
\text{(L1)}\; b \ne a\cdot x,\qquad
\text{(L2)}\; Q \ne P\cdot x,\qquad
\text{(L3)}\; P \ne Q\cdot x,\qquad
\text{(L4)}\; S \ne R\cdot x,\qquad
\text{(L5)}\; R \ne S\cdot x,
\]
for all $x$~\cite{zhang2005computer}.
In particular, (L1) implies $b\ne a$ (take $x=1$), but this also immediately
follows by looking at Wilkie's identity. Furthermore, Burris and Lee showed
that $a$ and $b$ cannot be integers \cite{burris1992}.
Thus, any witness elements are distinct
and lie outside the integer chain. Without loss of generality, we may therefore rename
the remaining domain so that $a=4$ and $b=5$, so $a$ and $b$ are two distinct non-integers.

Zhang also lists explicit ``integer-collapse'' conditions that must be avoided
by any countermodel with $W(a,b)$ false~\cite{zhang2005computer}.
Concretely, the following must all hold:
\begin{align*}
\text{(M01)}\;& 1+a \ne 1, & \text{(M02)}\;& 2+a \ne 1, &
\text{(M03)}\;& a+a \ne 1, \\
\text{(M04)}\;& a\cdot a \ne 1, & \text{(M05)}\;& 1+a\cdot a \ne 1, &
\text{(M06)}\;& a\cdot a\cdot a \ne 1, \\
\text{(M07)}\;& 1+a \ne a, & \text{(M08)}\;& 2+a \ne a, &
\text{(M09)}\;& a+a \ne a, \\
\text{(M10)}\;& a\cdot a \ne a, & \text{(M11)}\;& 1+a\cdot a \ne a, &
\text{(M12)}\;& 2+a \ne 1+a, \\
\text{(M13)}\;& a\cdot a \ne 1+a, &
\text{(M14)}\;& a\cdot a\cdot a \ne 1+a, &
\text{(M15)}\;& a\cdot a \ne 2+a, \\
\text{(M16)}\;& a\cdot a \ne a+a, &
\text{(M17)}\;& 1+a\cdot a \ne a\cdot a.
\end{align*}
These inequalities are encoded explicitly to prune the SAT search space.
Finally, the existence of a finite HSI-algebra of size $n$ is a finite
constraint satisfaction problem: the operation tables are finite and total and
all axioms are universal equations. This makes propositional satisfiability a
natural tool for the remaining open case $n=11$, and the next section gives the
complete SAT encoding.


\section{SAT Encoding}
\label{sec:sat}

We give a propositional encoding of the existence of a finite algebra
satisfying HSI while refuting Wilkie's identity.
The SAT problem asks whether a Boolean formula, usually given in conjunctive
normal form (CNF), has at least one truth assignment that satisfies all
clauses; for general background on SAT solving, see the
Handbook of Satisfiability~\cite{DBLP:series/faia/336}.
All equations are translated into Boolean formulas in conjunctive normal form
(CNF). Fix a finite domain
\[
D = \{1,2,\dots,n\}, \qquad n \ge 5,
\]
and two distinguished elements
\[
a := 4, \qquad b := 5.
\]
We introduce Boolean variables as follows. For each operation $\circ \in \{+,\times,\operatorname{exp}\}$ and all
$i,j,k \in D$ we introduce a Boolean variable $v^{\circ}_{i,j,k}$ with intended
meaning $v^{\circ}_{i,j,k} = \text{true} \iff i \circ j = k$.
Note that we use $\times$ instead of $\cdot$ in this section in order to improve
readability. For addition and
multiplication, commutativity is enforced by identifying
$v^{\circ}_{i,j,k} \equiv v^{\circ}_{j,i,k}$. For every composite term $t$
occurring in the encoding, we introduce variables $v_{t=k}$ ($k \in D$) with
intended meaning $v_{t=k} = \text{true} \iff t$ evaluates to $k$.

Each operation is required to be a total function. For all
$i,j \in D$ and $\circ \in \{+,\times,\operatorname{exp}\}$ we add
\[
\bigvee_{k \in D} v^{\circ}_{i,j,k}
\qquad\text{and}\qquad
\neg v^{\circ}_{i,j,k} \lor \neg v^{\circ}_{i,j,k'} \; (k \neq k').
\]
Together these clauses enforce that for each $i,j$ there is a unique
$k \in D$ with $i \circ j = k$.
We fix the canonical integers $1,2,3$ by the unit clauses
\[
v^{+}_{1,1,2} \qquad\text{and}\qquad v^{+}_{2,1,3},
\]
which remove isomorphic renamings of the integer chain; fixing $a=4$ and $b=5$
breaks additional domain symmetries.

The HSI axioms are encoded as universally quantified CNF clauses.
Commutativity of $+$ and $\times$ is handled by the variable indexing scheme, so
no additional clauses are required. Associativity for $+$ and $\times$ is
captured uniformly as follows: for all $x,y,z,u,v,w \in D$ and
$\circ \in \{+,\times\}$ we add
\[
\neg v^{\circ}_{x,y,u}
\lor
\neg v^{\circ}_{u,z,w}
\lor
\neg v^{\circ}_{y,z,v}
\lor
v^{\circ}_{x,v,w}.
\]
Distributivity is encoded by clauses, for all $x,y,z,u,p,q,w \in D$,
\[
\neg v^{+}_{y,z,u}
\lor
\neg v^{\times}_{x,u,w}
\lor
\neg v^{\times}_{x,y,p}
\lor
\neg v^{\times}_{x,z,q}
\lor
v^{+}_{p,q,w},
\]
corresponding to $x \times (y+z) = (x\times y) + (x\times z)$. The exponentiation
laws are encoded, for all $x,y,z,u,p,q,w \in D$, by
\[
\neg v^{+}_{y,z,u}
\lor
\neg v^{\operatorname{exp}}_{x,u,w}
\lor
\neg v^{\operatorname{exp}}_{x,y,p}
\lor
\neg v^{\operatorname{exp}}_{x,z,q}
\lor
v^{\times}_{p,q,w},
\]
\[
\neg v^{\times}_{x,y,u}
\lor
\neg v^{\operatorname{exp}}_{u,z,w}
\lor
\neg v^{\operatorname{exp}}_{x,z,p}
\lor
\neg v^{\operatorname{exp}}_{y,z,q}
\lor
v^{\times}_{p,q,w},
\]
and
\[
\neg v^{\times}_{y,z,u}
\lor
\neg v^{\operatorname{exp}}_{x,u,w}
\lor
\neg v^{\operatorname{exp}}_{x,y,p}
\lor
v^{\operatorname{exp}}_{p,z,w},
\]
corresponding to $x^{y+z} = x^y \times x^z$, $(x\times y)^z = x^z \times y^z$, and
$(x^y)^z = x^{y\times z}$. The identity laws are enforced by the unit clauses,
for all $x \in D$,
\[
v^{\times}_{x,1,x}, \qquad v^{\operatorname{exp}}_{1,x,1}, \qquad
v^{\operatorname{exp}}_{x,1,x}.
\]

Composite terms are handled recursively. A composite term $t=(\mathsf{op},t_1,t_2)$ is encoded by a
one-hot constraint
\[
\bigvee_{k \in D} v_{t=k}
\qquad\text{and}\qquad
\neg v_{t=k} \lor \neg v_{t=k'} \; (k \neq k'),
\]
and by forward evaluation clauses, for all $i,j,k \in D$,
\[
\neg v_{t_1=i} \lor \neg v_{t_2=j} \lor \neg v^{\circ}_{i,j,k} \lor v_{t=k},
\]
encoding $(t_1=i \land t_2=j \land i\circ j=k) \Rightarrow (t=k)$.

To refute Wilkie's identity we encode all subterms recursively and fix the
witness pair $(a,b)=(4,5)$ to encode the instance $W(a,b)$. Define the base terms
\[
\begin{aligned}
a^2 &= a\times a, & P &= 1 + a, & Q &= P + a^2, \\
a^3 &= a^2\times a, & R &= 1 + a^3, & a^4 &= a^2\times a^2, &
S &= (1 + a^2) + a^4,
\end{aligned}
\]
then
\[
\begin{aligned}
T_1 &= P^a + Q^a, & T_2 &= R^b + S^b, &
\mathrm{LHS} &= (T_1)^b \times (T_2)^a, \\
U_1 &= P^b + Q^b, & U_2 &= R^a + S^a, &
\mathrm{RHS} &= (U_1)^a \times (U_2)^b.
\end{aligned}
\]
We enforce $\mathrm{LHS} \ne \mathrm{RHS}$ by the clauses, for all $k \in D$,
\[
\neg v_{\mathrm{LHS}=k} \lor \neg v_{\mathrm{RHS}=k}.
\]
We also add the divisibility lemmas (L1--L5): for all $x,i,j \in D$,
\[
\neg v_{P=i} \lor \neg v^{\times}_{i,x,j} \lor \neg v_{Q=j},
\]
encodes L2; clauses for L1 and L3--L5 are obtained analogously, thereby
excluding divisibility relations that would force Wilkie's identity to hold.
The integer-collapse lemmas (M01--M17) encode each forbidden equality $t=u$ as
\[
\bigwedge_{k \in D}\;(\neg v_{t=k} \lor \neg v_{u=k}).
\]
For the lemmas that mention $a^2$ or $a^3$ (M05, M06, M11, M14, M17), we
guard the inequality by the value of that subterm; for example, if $a^2=k$ then
we forbid $t$ and $u$ from both taking $k$, yielding
\[
\bigwedge_{k \in D}\;(\neg v_{a^2=k} \lor \neg v_{t=k} \lor \neg v_{u=k})
\]
and analogously with $a^3$ when it appears.
The resulting CNF formula is satisfiable if and only if there exists an
$n$-element HSI-algebra in which Wilkie's identity fails.

For the commutative operations $+$ and $\times$, variables are allocated only for
pairs with $i \le j$ (the $n$ diagonal pairs plus $\binom{n}{2}$ off-diagonals),
yielding $n^2(n+1)/2$ variables per operation and $n^2(n+1)$ in
total. Exponentiation contributes $n^3$ variables, and the encoder introduces
one-hot evaluation variables for each of the $26$ composite terms in the Wilkie
expression and pruning lemmas, yielding $26n$ more variables. Thus
\[
V(n)=n^2(n+1) + n^3 + 26n = 2n^3 + n^2 + 26n.
\]
Each operation entry is constrained by an \emph{exactly-one} encoding with one
$n$-literal clause and $\binom{n}{2}$ binary clauses. There are $n(n+1)/2$
entries for each of $+$ and $\times$, and $n^2$ entries for exponentiation,
so the number of functionality clauses is
\[
F(n)=\Big(2\times \frac{n(n+1)}{2} + n^2\Big)\!\Big(1+\frac{n(n-1)}{2}\Big)
      =\frac{(2n^2+n)(n^2-n+2)}{2}.
\]
The HSI axioms contribute $2n^6$ clauses for associativity, $n^7$ clauses for
distributivity, and $n^7$, $n^7$, and $n^6$ clauses for the three
exponentiation laws, plus $3n$ identity-unit clauses, giving
\[
H(n)=3n^7 + 3n^6 + 3n.
\]
Each composite term is encoded once (memoized), adding a one-hot constraint
with $1+\binom{n}{2}$ clauses. Forward evaluation clauses contribute
$9n^3 + 15n^2 + 2n$ clauses for the Wilkie term DAG, and the $26$ one-hot
constraints with $1+\binom{n}{2}$ clauses each bring this total to
\[
T(n)=9n^3 + 28n^2 - 11n + 26.
\]
Let $I(n)=n$ denote the number of clauses enforcing the inequality
$\mathrm{LHS} \ne \mathrm{RHS}$.
The divisibility lemmas add $n$ clauses for L1 and $4n^3$ clauses for
L2--L5, and the integer-collapse lemmas contribute $n^2 + 8n + 8$, so
overall the number of clauses originating from the pruning lemmas is
\[
P(n)=4n^3 + n^2 + 9n + 8.
\]
The canonical constraints for the witness pair add $U(n)=2$ unit clauses.
Summing all components gives
\begin{align*}
C(n) &= F(n)+H(n)+T(n)+I(n)+P(n)+U(n) \\
     &= 3n^7 + 3n^6 + n^4 + \frac{25}{2}n^3 + \frac{61}{2}n^2 + 3n + 36.
\end{align*}
Thus $V(n)=\Theta(n^3)$ and $C(n)=\Theta(n^7)$.
Table~\ref{tab:sizes} reports representative instance sizes computed from
these formulas.

\begin{table}[h]
\centering
\begin{tabular}{rrr}
\hline
$n$ & Variables & Clauses \\
\hline
5 & 405 & 284{,}251 \\
6 & 624 & 984{,}924 \\
7 & 917 & 2{,}831{,}816 \\
8 & 1{,}296 & 7{,}090{,}396 \\
9 & 1{,}773 & 15{,}961{,}437 \\
10 & 2{,}360 & 33{,}025{,}616 \\
11 & 3{,}069 & 63{,}811{,}234 \\
\hline
\end{tabular}
\caption{Instance sizes for generated CNF files}
\label{tab:sizes}
\end{table}


\section{Experimental Results}
\label{sec:experiments}

We solved the generated CNF instances for $n=5,\dots,11$ using the clause-portfolio
solvers Gimsatul and MallobSat~\cite{fleuryBiere2022gimsatul,schreiber2024mallobsat}.
Both are clause-portfolio solvers, running diverse strategies in parallel and
sharing learned clauses across workers.
All instances for $n=5,\dots,11$ were found to be unsatisfiable.
All experiments were run on a machine with 64 Intel Xeon Gold 6526Y CPUs
(2.8\,GHz base frequency) and 300\,GB RAM.
As a single-threaded baseline, we also ran CaDiCaL~\cite{cadical2024}; for $n=11$
it timed out after 4 days.

In order to verify the unsatisfiability results obtained by the SAT solvers,
we checked the results independently by certifying proof outputs.
Gimsatul was configured to emit DRAT proofs. We verified the Gimsatul proofs for
$n=5$ through $n=10$ with the formally verified GRAT toolchain (\texttt{gratgen}/\texttt{gratchk})~\cite{lammich2020grat}.
For $n=11$, we attempted \texttt{drat-trim}~\cite{drattrim}; projected runtimes
exceeded 50 days on our hardware, and the GRAT verification ran out of memory.
All runs used 48 threads, except for MallobSat at $n=11$, which was reduced to
40 threads due to proof-production issues.

MallobSat was configured to emit LRUP proofs~\cite{michaelson2025producing}. LRUP (and its backward-compatible
extension LRAT) provides explicit checking hints, making verification fast
~\cite{cruzfilipe2016lrat}. We verified all MallobSat proofs for
$n=5,\dots,11$ using \texttt{lrat\_isa}, a formally verified checker~\cite{lammich2024lratisa}.
In contrast to Gimsatul/GRAT, checking the certificate needed less time and
memory than SAT-solving.
For $n=11$, \texttt{lrat\_isa} completed in 50.0\,min.

Table~\ref{tab:solve-times} compares wall-clock solve times. Gimsatul is
fastest for $n=5$--$7$ and $n=9$, while MallobSat is slightly faster at $n=8$ and
becomes faster at $n=10$ (about $1.50\times$) and substantially faster at $n=11$
(about $2.77\times$). The single-threaded CaDiCaL baseline is competitive for
$n\le 7$, but slows down by $n=8$ and $n=9$; at $n=10$ it is $11.1\times$ slower
than MallobSat and $7.41\times$ slower than Gimsatul, and it times out at $n=11$.

As a sanity check, we also tried whether our encoding was satisfiable for $n=12$.
This produced the 12-element algebra shown in Table~\ref{tab:operations-12}
which is not isomorphic to the previously known
12-element model presented in~\cite{burris2005saga}. The satisfiable instance was
found by Gimsatul with 16 threads in 47.4\,min. We validated it with a small
checker that reconstructs the operation tables from the SAT assignment and then
evaluates all HSI axioms and Wilkie's identity on the resulting algebra.
Non-isomorphism was verified by exhaustive
permutation search, checking whether any permutation simultaneously preserves
$+$, $\times$, and $\operatorname{exp}$.

We also tried to determine the total number of 12-element GBAs,
but could not finish the task within two weeks on our hardware.
To that end, we ran Gimsatul (non-incrementally) by extending our
encoding to block previously found algebras. After each iteration, we validated
that the algebra is a GBA and checked whether it is isomorphic to a previously
found one. That way, we found more than 500 non-isomorphic 12-element GBAs.

\begin{table}[t]
\centering
\begin{tabular}{llll}
\hline
Solver & Proof format & Checker & Verified $n$ \\
\hline
Gimsatul & DRAT & \texttt{gratgen}/\texttt{gratchk} & $5$--$10$ \\
MallobSat & LRUP & \texttt{lrat\_isa} & $5$--$11$ \\
\hline
\end{tabular}
\caption{Proof production and verification summary.}
\label{tab:proofs}
\end{table}

\begin{table}[t]
\centering
\begin{tabular}{llll}
\hline
$n$ & Gimsatul time & MallobSat time & CaDiCaL time \\
\hline
5 & 0.16\,s & 1.31\,s & 0.10\,s \\
6 & 1.39\,s & 2.01\,s & 2.23\,s \\
7 & 4.35\,s & 5.29\,s & 5.46\,s \\
8 & 15.32\,s & 14.28\,s & 24.62\,s \\
9 & 81.00\,s & 95.18\,s & 5.40\,min \\
10 & 25.2\,min & 16.8\,min & 3.11\,h \\
11 & 21.0\,h & 7.59\,h & 4\,d (timeout) \\
\hline
\end{tabular}
\caption{Wall-clock solve times on the benchmark machine}
\label{tab:solve-times}
\end{table}

\section{Conclusion}
\label{sec:conclusion}
We established that there is no HSI countermodel to Wilkie’s identity of size
$n \le 11$ by SAT solving with independently verified UNSAT certificates. This
settles the last open size below 12 and confirms that 12 is the smallest size
where countermodels can occur. As a byproduct, we found 12-element
countermodels which are not isomorphic to the previously known one.
The scripts and files to reproduce our results can be found on Zenodo.%
\footnote{\url{https://doi.org/10.5281/zenodo.18568303}}

\bibliography{main}

\end{document}